\documentclass[12pt,a4paper]{article}
\usepackage[utf8]{inputenc}
\usepackage[english]{babel}
\usepackage{amsmath}
\usepackage{amsfonts}
\usepackage{amssymb}
\usepackage{makeidx}
\usepackage{graphicx}
\usepackage[left=2cm,right=2cm,top=2cm,bottom=2cm]{geometry}
\usepackage{physics}
\def\a{\alpha}

\def\b{\beta}

\def\m{\mu}

\usepackage{authblk}

\title{More than one Author with different Affiliations}
\author[1]{Jitendra Pal\thanks{jeetupal007@gmail.com}}
\author[1]{Arnab Mukherjee\thanks{mukherji.arn@gmail.com}}
\author[2]{Arindam Lala\thanks{arindam.physics1@gmail.com, arindam.lala@pucv.cl}}
\author[1]{Dibakar Roychowdhury\thanks{dibakarphys@gmail.com, dibakarfph@iitr.ac.in}}
\affil[1]{Department of Physics, Indian Institute of Technology Roorkee, Roorkee 247667 Uttarakhand, India}
\affil[2]{Instituto de F\'{i}sica, Pontificia Universidad Cat\'{o}lica de Valpara\'{i}so, Casilla 4059, Valparaiso, Chile}

\begin{document}
\date{}
%%%%%%%%%%%%%%%%%%%%
\title{{\bf{\Large Analytic (non)integrability of Arutyunov-Bassi-Lacroix model}}}
%%%%%%%%%%%%%%%%%%%%
\iffalse
\author{{\bf {\normalsize Arindam Lala}$
$\thanks{E-mail:  arindam.physics1@gmail.com, arindam.lala@pucv.cl}}\\
 {\normalsize  Instituto de F\'{i}sica, Pontificia Universidad Cat\'{o}lica de Valpara\'{i}so,}\\
  {\normalsize Casilla 4059, Valparaiso, CHILE,}
\\[0.3cm]
 {\bf {\normalsize Dibakar Roychowdhury}$
$\thanks{E-mail:  dibakarphys@gmail.com, dibakarfph@iitr.ac.in}}\\
 {\normalsize  Department of Physics, Indian Institute of Technology Roorkee,}\\
  {\normalsize Roorkee 247667 Uttarakhand, India}
\\[0.3cm]
}
\fi
%\date{}
%%%%%%%%%%%%%%%%%%%%
\maketitle
%%%%%%%%%%%%%%%%%%%%
\begin{abstract}
We use the notion of the gauge/string duality and discuss the Liouvillian (non) integrability criteria for string
sigma models in the context of recently proposed Arutyunov-Bassi-Lacroix (ABL) model [JHEP \textbf{03}
(2021), 062]. Our analysis complements those previous results due to numerical analysis as well as
Lax pair formulation. We consider a winding string ansatz for the deformed torus $T^{\qty(\lambda_{1},\lambda_{2},
\lambda)}_{k}$ which can be interpreted as a system of coupled pendulums. Our analysis reveals the Liouvillian
nonintegrablity of the associated sigma model. We also obtain the \emph{generalized} decoupling limit and confirm
the analytic integrability for the decoupled sector.

\end{abstract}
%%%%%%%%%%%%%%%%%%%%
\section{Introduction and motivation}\label{intro}
Integrable systems are very special - they posses an infinite tower of conserved charges those are in involution.
The notion of integrability was first introduced in the seminal work of Bethe in an attempt to solve the Heisenberg
spin chain model \cite{Bethe:1931hc}. Subsequently, these techniques were further developed and applied to
models like QCD in \cite{Lipatov:1993yb}. In recent years, following the pioneering work of \cite{Minahan:2002ve},
there has been a surge in the study of integrable systems within the framework of gauge/string duality 
\cite{Frolov:1999}-\cite{Rigatos:2020igd}.\medskip

In this work, we use the methodology of the AdS/CFT duality \cite{Maldacena:1997re,Witten:1998qj,Aharony:1999ti}
in order to study the dynamics of string solitons over Arutyunov-Bassi-Lacroix (ABL) background $\mathcal{R}
\times \mathcal{T}^{\qty(\lambda_{1},\lambda_{2},\lambda)}_{k}$ \cite{Arutyunov:2020sdo,Ishii:2021asw}. Along the
way, we explore the classical (non)integrability of the associated phase space. This ABL background can be considered
as a generalization of the Einsteinian $T^{1,1}$ manifold \cite{Klebanov:1998hh} and is given by 
\cite{Arutyunov:2020sdo,Ishii:2021asw}
%%%%
\begin{align}
\dd s^{2} &=- \dd t^2+\sum_{i=1}^{2} \lambda_{i}^{2}\left(\dd \theta_{i}^{2}+
\sin ^{2} \theta_{i} \dd \phi_{i}^{2}\right)+\lambda^{2}\left(\dd \psi+\cos \theta_{1}
\dd \phi_{1}+\cos \theta_{2} \dd \phi_{2}\right)^{2} \, , \label{ABL:metric} \\
B_{2} &= k\left(\dd \psi+\cos \theta_{1} \dd \phi_{1}\right) \wedge\left(\dd \psi
+\cos \theta_{2} \dd \phi_{2}\right) \,.  \label{B:field}
\end{align}
%%%%
Here $\lambda_{1}$, $\lambda_{2}$, $\lambda$ and $k$ are real parameters. Interestingly, the classical string dynamics
is integrable only when $k=\lambda^{2}$. When $k\neq \lambda^{2}$, the integrability of the system is lost and it becomes
chaotic. This has been confirmed very recently using numerical techniques \cite{Ishii:2021asw}.\medskip

In this paper, we complement all those previous results \cite{Arutyunov:2020sdo,Ishii:2021asw} by taking a third path
which is based on the notion of Kovacic's algorithm \cite{Kovacic:1986,Saunders:1981}. The algorithm essentially offers
a set of rules in order to verify the Liouvillian non-integrability criteria for a classical 2$d$ sigma model over general
backgrounds. As our analysis reveals, it establishes a clear compatibility between the Liouvillian (non)integrability and
the Lax pair formulation of 2$d$ sigma model over ABL background.\medskip

In our analysis, we consider winding string ansatz where a string is wrapped around all the isometry directions
($\phi_{1},\phi_{2},\psi$). We observe that when the modes of the string along the ($\phi_{1},\phi_{2}$) directions are
coupled, the dynamics of the system can be modelled as a system of coupled pendulums. Subsequently, the application
of the Kovacic's algorithm ensures non-integrability of this coupled pendulum model. We also observe a \emph{generalized}
special limit when these pendulums decouple and become integrable. Interestingly, the decoupling limit studied in 
\cite{Arutyunov:2020sdo,Ishii:2021asw} can be seen as a special case of our analysis. While achieving this later limit, we
observe that the string must satisfy certain conditions that have been discussed in detail in the section \ref{decoup}.
\medskip

The structure of the paper is as follows: In section \ref{abl:sigma} we study the string sigma model in the ABL
background (\ref{ABL:metric}). Using winding string ansatz, we explicitly compute the set of differential equations
that represents two coupled harmonic oscillators. We explicitly determined the conditions that make the Virasoro
constraints to satisfy. Furthermore, we apply the Kovacic's algorithm to show analytically the non-integrability of this
system of coupled pendulums. In section \ref{decoup} we analyze the decoupling limits of these coupled pendulum
system and checked the integrability in individual system. Finally, we conclude in section \ref{conclusion}. Additionally,
we give a brief account of the Kovacic's algorithm in Appendix \ref{kovacic} in order to make the article self-contained.
Appendix \ref{coefschi} contains lengthy expressions of certain constant coefficients.

%At this point we notice that non-integrability of string sigma model in the $T^{1,1}$ manifold has been studied analytic as
%well as numerically in \cite{Basu:2011b,Basu:2011fw}. 

%%%%%%%%%%%%%%%%%%%%
\section{Classical sigma model on $\mathcal{R}\times T^{\lambda_{2},\lambda_{1},\lambda}_{k}$}\label{abl:sigma}
The classical $2d$ string sigma model can be expressed in the conformal gauge as \cite{Tseytlin:1988rr}
%%%%
\begin{equation}\label{Act::sigma}
S_{P} = \frac{1}{4\pi\alpha'} \int \dd \tau \dd \sigma \qty(\eta^{\a\b}G_{MN}
+\epsilon^{\a\b}B_{MN}) \partial_{\a}X^{M} \partial_{\b}X^{N} = \frac{1}{4\pi\alpha'}
\int \dd\tau \dd\sigma L_{P} \, .
\end{equation}
%%%%
Here $\a,\b$ are string worldsheet coordinates with metric $\eta_{\a\b}=\text{diag}\qty(-1,1)$, and the target
space coordinates are labelled as $\displaystyle X^{\m}=\qty(t,\theta_{1},\theta_{2},\phi_{1},\phi_{2},\psi)$.

Using (\ref{ABL:metric}) and (\ref{B:field}), the Lagrangian density $L_{P}$ appearing in (\ref{Act::sigma}) can
be written as
%%%%
\begin{align}\label{Ldensity:ABL}
\begin{split}
L_{P} &= \qty(\dot{t}^{2} - t'^{2}) + \lambda_{1}^{2} \qty(\theta_{1}'^{2} - \dot{\theta_{1}}^{2})
+ \lambda_{2}^{2} \qty(\theta_{2}'^{2} - \dot{\theta_{2}}^{2}) + \lambda^{2} \qty(\psi'^{2} -
\dot{\psi}^{2}) + \qty(\lambda_{1}^{2} \sin^{2}\theta_{1} + \lambda^{2} \cos^{2}\theta_{1}) \\
& \qquad \qty(\phi_{1}'^{2} - \dot{\phi_{1}}^{2}) + \qty(\lambda_{2}^{2} \sin^{2}\theta_{2} +
\lambda^{2} \cos^{2}\theta_{2}) \qty(\phi_{2}'^{2} - \dot{\phi_{2}}^{2}) + 2\lambda^{2}\cos
\theta_{1} \qty(\psi' \phi_{1}' - \dot{\psi}\dot{\phi}_{1})  \\
& \qquad + 2\lambda^{2}\cos\theta_{2} \qty(\psi' \phi_{2}' - \dot{\psi}\dot{\phi}_{2}) + 2
\lambda^{2} \cos\theta_{1} \cos\theta_{2} \qty(\phi_{1}' \phi_{2}' - \dot{\phi}_{1}\dot{\phi}_{2})
+2 k \cos\theta_{1} \qty(\dot{\phi}_{1} \psi' - \dot{\psi}\phi_{1}')   \\
& \qquad + 2 k \cos\theta_{2} \qty(\dot{\psi} \phi_{2}' - \dot{\phi}_{2}\psi') + 2k \cos\theta_{1}
\cos\theta_{2} \qty(\dot{\phi}_{1} \phi_{2}' - \dot{\phi}_{2}\phi_{1}') \, .
\end{split}
\end{align}
%%%%

The conserved charges associated with the string $\sigma$-model (\ref{Act::sigma}) are simply derived from the
above density (\ref{Ldensity:ABL}) and are written as
%%%%
\begin{eqnarray}
E &=& \frac{\partial L_{P}}{\partial \dot{t}} = 2\dot{t} \, ,    \label{con:energy}\\[5pt]
P_{\psi} &=& \frac{\partial L_{P}}{\partial \dot{\psi}} = -2 \qty[ \lambda^{2}\dot{\psi} + \qty(k
\phi_{1}' + \lambda^{2} \dot{\phi_{1}})\cos\theta_{1} - \qty(k\phi_{2}' - \lambda^{2}
\dot{\phi_{2}})\cos\theta_{2}] \, ,  \label{con:mompsi}\\[5pt]
P_{\phi_{1}} &=& \frac{\partial L_{P}}{\partial\dot{\phi_{1}}} = -2 \big[\qty(\lambda_{1}^{2}
\sin^{2}\theta_{1} + \lambda^{2} \cos^{2}\theta_{1}) \dot{\phi_{1}} - \qty(k\psi' - \lambda^{2}
\dot{\psi})\cos\theta_{1} \\ \nonumber
&& \qquad \qquad - \qty(k \phi_{2}' - \lambda^{2} \dot{\phi_{2}})\cos\theta_{1}\cos\theta_{2}
\big] \, ,  \label{con:momphi1}\\[5pt]
P_{\phi_{2}} &=& \frac{\partial L_{P}}{\partial\dot{\phi_{1}}} = -2 \big[\qty(\lambda_{2}^{2}
\sin^{2}\theta_{2} + \lambda^{2} \cos^{2}\theta_{2}) \dot{\phi_{2}} + \qty(k\psi' + \lambda^{2}
\dot{\psi})\cos\theta_{2} \\ \nonumber
&& \qquad \qquad + \qty(k \phi_{1}' + \lambda^{2} \dot{\phi_{1}})\cos\theta_{1}\cos\theta_{2}
\big] \, . \label{con:momphi2}
\end{eqnarray}
%%%%

Next, we consider the winding string ansatz of the following form \cite{Ishii:2021asw}
%%%%
\begin{align}
\begin{split}\label{ansatz::1a}
t ={}& \tau \, ,  \qquad\qquad\qquad   \theta_{1} =\theta_{1}\qty(\tau) \, , \qquad\qquad\qquad
   \theta_{2} =\theta_{2}\qty(\tau) \, , \end{split} \\[8pt]
\begin{split}\label{ansatz::1b}
\phi_{1} ={}& \omega_{1}\tau + \ell_{1} \sigma \, ,  \qquad  \phi_{2} = \omega_{2}\tau + \ell_{2} \sigma \, ,
\qquad\qquad  \psi = \alpha \tau + \ell_{3}\sigma \,,
\end{split}
\end{align}
%%%%
where $\ell_{i}$s are the respective winding numbers along the isometries of the target space.

Substituting (\ref{ansatz::1a}) and (\ref{ansatz::1b}) into (\ref{Ldensity:ABL}) we find
%%%%
\begin{align}\label{Lag::1D}
\begin{split}
\tilde{L}_{P} &= 1-\lambda_{1}^{2} \dot{\theta}_{1}^{2} -\lambda_{2}^{2} \dot{\theta}_{2}^{2}
+\lambda^{2} \qty(\ell_{3}^{2}-\alpha^{2}) + \qty(\lambda_{1}^{2} \sin^{2}\theta_{1}+
\lambda^{2} \cos^{2}\theta_{1}) \qty(\ell_{1}^{2}-\omega_{1}^{2})  \\[3pt]
& \quad + \qty(\lambda_{1}^{2}\sin^{2}\theta_{2}+\lambda^{2} \cos^{2}\theta_{2})
\qty(\ell_{2}^{2}-\omega_{2}^{2}) + 2\lambda^{2} \cos\theta_{1}\qty(\ell_{1}\ell_{3}-
\alpha \omega_{1}) \\[3pt]
& \quad + 2\lambda^{2}\cos\theta_{2} \qty(\ell_{2}\ell_{3}-\alpha \omega_{2}) + 2
\lambda^{2}\cos\theta_{1} \cos\theta_{2} \qty(\ell_{1}\ell_{2}-\omega_{1}\omega_{2})
\\[3pt]
&\quad +2 k \cos\theta_{1} \qty(\omega_{1}\ell_{3}-\alpha\ell_{1})
+2 k \cos\theta_{2}\qty(\alpha\ell_{2}-\omega_{2}\ell_{3}) + 2 k \cos\theta_{1} \cos\theta_{2}
\qty(\omega_{1}\ell_{2}-\omega_{2}\ell_{1}).
\end{split}
\end{align}
%%%%

The equations of motion corresponding to $\theta_{i}$ ($i=1,2$) can be derived from the above Lagrangian density
(\ref{Lag::1D}) as
%%%%
\begin{eqnarray}
%\begin{split}
\ddot{\theta}_{1} & +\qty(1-\frac{\lambda^{2}}{\lambda_{1}^{2}})\qty(\ell_{1}^{2}-\omega_{1}^{2})
\sin\theta_{1} \cos\theta_{1} - \frac{\sin\theta_{1}}{\lambda_{1}^{2}}\qty \Big[ k \qty(\omega_{1}\ell_{3}
-\alpha\ell_{1})+\lambda^{2} \qty(\ell_{1}\ell_{3}-\alpha \omega_{1}) ]  \nonumber \\[3pt] 
& \qquad - \frac{\sin\theta_{1}\cos\theta_{2}}{\lambda_{1}^{2}} \qty \Big[ k \qty(\omega_{1}\ell_{2}
-\omega_{2}\ell_{1})+\lambda^{2} \qty(\ell_{1}\ell_{2}-\omega_{1} \omega_{2}) ] = 0 \, ,  
\label{eom:theta1} \\[3pt]
\ddot{\theta}_{2} & +\qty(1-\frac{\lambda^{2}}{\lambda_{2}^{2}})\qty(\ell_{2}^{2}-\omega_{2}^{2})
\sin\theta_{2} \cos\theta_{2} - \frac{\sin\theta_{2}}{\lambda_{2}^{2}}\qty \Big[ k \qty(\alpha\ell_{2}
-\omega_{2}\ell_{3})+\lambda^{2} \qty(\ell_{2}\ell_{3}-\alpha \omega_{2}) ]  \nonumber \\[3pt] 
& \qquad - \frac{\sin\theta_{2}\cos\theta_{1}}{\lambda_{2}^{2}} \qty \Big[ k \qty(\omega_{1}\ell_{2}
-\omega_{2}\ell_{1})+\lambda^{2} \qty(\ell_{1}\ell_{2}-\omega_{1} \omega_{2}) ] = 0 \, .
\label{eom:theta2}
%\end{split}
\end{eqnarray}
%%%%

This looks like an interacting oscillator model. In other words, the sigma model could be thought of as a model
of coupled oscillators which are mutually interacting.

In addition, the corresponding Virasoro constraints can be written as
%%%%
\begin{align}\label{Ttt}
\begin{split}
T_{\tau\tau} = T_{\sigma\sigma} &= \frac{1}{2} \Big[ -1+\lambda_{1}^{2}\dot{\theta_{1}^{2}}
+\lambda_{2}^{2} \dot{\theta}_{2}^{2}+\lambda^{2} \qty(\ell_{3}^{2}+\alpha^{2}) +
\qty(\lambda_{1}^{2} \sin^{2}\theta_{1}+\lambda^{2} \cos^{2}\theta_{1}) \qty(\ell_{1}^{2}
+\omega_{1}^{2})  \\[3pt]
& \quad + \qty(\lambda_{1}^{2}\sin^{2}\theta_{2}+\lambda^{2} \cos^{2}\theta_{2})
\qty(\ell_{2}^{2}+\omega_{2}^{2}) + 2\lambda^{2} \cos\theta_{1}\qty(\ell_{1}\ell_{3}+
\alpha \omega_{1}) \\[3pt]
& \quad + 2\lambda^{2}\cos\theta_{2} \qty(\ell_{2}\ell_{3}+\alpha \omega_{2}) + 2
\lambda^{2}\cos\theta_{1} \cos\theta_{2} \qty(\ell_{1}\ell_{2} + \omega_{1}\omega_{2})
\Big] \, ,
\end{split}
\end{align}
%%%%

%%%%
\begin{align}\label{Tts}
\begin{split}
T_{\tau\sigma} = T_{\sigma\tau} &= \lambda^{2}\alpha\ell_{3} + \qty(\lambda_{1}^{2} \sin^{2}
\theta_{1}+\lambda^{2} \cos^{2}\theta_{1}) \omega_{1}\ell_{1} + \qty(\lambda_{1}^{2}\sin^{2}
\theta_{2}+\lambda^{2}\cos^{2}\theta_{2}) \omega_{2}\ell_{2}   \\[3pt]
& \quad + \lambda^{2} \cos\theta_{1} \qty(\alpha\ell_{1}+\omega_{1}\ell_{3}) + \lambda^{2}
\cos\theta_{2} \qty(\alpha\ell_{2}+\omega_{2}\ell_{3})   \\[3pt]
& \quad + \lambda^{2}\cos\theta_{1} \cos\theta_{2}\qty(\omega_{1}\ell_{2} + \ell_{1} \omega_{2})
\Big] \, .
\end{split}
\end{align}
%%%%

\subsection{Consistency requirements}\label{consistency}
Using the equations of motion (\ref{eom:theta1}), (\ref{eom:theta2})  we may write
%%%%
\begin{align}\label{der:Ttt}
\begin{split}
&\partial_{\tau} T_{\tau\tau}  \\[3pt]
& =\omega_{1}^{2}\qty(\lambda_{1}^{2}-\lambda^{2})
\sin 2\theta_{1} \, \dot{\theta_{1}} + \omega_{2}^{2}\qty(\lambda_{2}^{2}-\lambda^{2})
\sin 2\theta_{2} \, \dot{\theta_{2}} - 2\alpha\lambda^{2} \qty(\omega_{1}\sin \theta_{1} \,
\dot{\theta_{1}} + \omega_{2}\sin \theta_{2} \, \dot{\theta_{2}} )   \\[3pt]
&	\quad - 2\lambda^{2}\omega_{1}\omega_{2} \qty( \sin \theta_{1} \cos \theta_{2}
\, \dot{\theta_{1}} + \sin \theta_{2} \cos \theta_{1} \, \dot{\theta_{2}} ) + k \sin \theta_{1} \,
\dot{\theta_{1}} \Big[ \omega_{1} \qty( \ell_{3} + \ell_{2} \cos\theta_{2})  \\[3pt]
& \quad - \ell_{1} \qty( \alpha + \omega_{2} \cos\theta_{2}) \Big] + k \sin \theta_{2} \,
\dot{\theta_{2}} \Big[ \ell_{2} \qty( \alpha + \omega_{1} \cos\theta_{1}) - \omega_{2} \qty(
\ell_{3} + \ell_{1} \cos\theta_{1}) \Big] \, ,
\end{split}
\end{align}
%%%%

%%%%
\begin{align}\label{der:Tts}
\begin{split}
\partial_{\tau} T_{\tau\sigma}
& = 2 \omega_{1}\ell_{1} \sin\theta_{1}\cos\theta_{1}\cdot \dot{\theta}_{1}\qty( \lambda_{1}^{2}
-\lambda^{2} ) + 2 \omega_{2}\ell_{2} \sin\theta_{2}\cos\theta_{2}\cdot \dot{\theta}_{2}
\qty( \lambda_{2}^{2}-\lambda^{2} )  \\[3pt]
& \quad - \lambda^{2} \qty(\alpha\ell_{1}+\omega_{1}\ell_{3})
\sin\theta_{1}\cdot \dot{\theta}_{1} - \lambda^{2} \qty(\alpha\ell_{2}+\omega_{2}\ell_{3})
\sin\theta_{2}\cdot \dot{\theta}_{2}  \\[3pt]
& \quad - \lambda^{2} \qty(\omega_{1}\ell_{2}+\omega_{2}\ell_{1})
\qty(\sin\theta_{1}\cos\theta_{2}\cdot \dot{\theta}_{1} + \sin\theta_{2}\cos\theta_{1}\cdot 
\dot{\theta}_{2})  \, .
\end{split}
\end{align}
%%%%

In the next step we demand the invariance of the conserved charges $\mathcal{Q}_{i}$ associated with the
reduced sigma model (\ref{Lag::1D}). This allows us to compute the following constraint equations\footnote{
Here we have used the following standard definition of the conserved charge: $\displaystyle J_{i} =
\frac{1}{2\pi\alpha'} \int_{0}^{2\pi} \dd\sigma P_{i}$, $i=\qty( \psi,\phi_{1},\phi_{2} )$ and the $P_{i}$s are given
in (\ref{con:mompsi})-(\ref{con:momphi2}).}:
%%%%
\begin{eqnarray}
\partial_{\tau}J_{\psi} &=& \qty(k \ell_{1} + \lambda^{2}\omega_{1}) \sin\theta_{1}
\cdot \dot{\theta}_{1} - \qty(k \ell_{2} - \lambda^{2}\omega_{2}) \sin\theta_{2}
\cdot \dot{\theta}_{2} = 0 \, , \label{con:psi}  \\[7pt]
\partial_{\tau}J_{\phi_{1}} &=& \qty[2\omega_{1} \qty(\lambda_{1}^{2} - \lambda^{2})\sin\theta_{1}
\cos\theta_{1} + \sin\theta_{1} \qty(k\ell_{3} - \alpha \lambda^{2}) + \sin\theta_{1}\cos\theta_{2}
\qty(k\ell_{2} - \omega_{2}\lambda^{2}) ] \cdot \dot{\theta}_{1} \nonumber \\[4pt]
&& + \qty(k\ell_{2} - \omega_{2} \lambda^{2}) \sin\theta_{2}\cos\theta_{1} \cdot \dot{\theta}_{2}
=0  \, ,  \label{con:phi1}   \\[7pt]
\partial_{\tau}J_{\phi_{2}} &=& \qty[2\omega_{2} \qty(\lambda_{2}^{2} - \lambda^{2})\sin\theta_{2}
\cos\theta_{2} - \sin\theta_{2} \qty(k\ell_{3} + \alpha \lambda^{2}) - \sin\theta_{2}\cos\theta_{1}
\qty(k\ell_{1} + \omega_{1}\lambda^{2}) ] \cdot \dot{\theta}_{2} \nonumber \\[4pt] 
&& - \qty(k\ell_{1} + \omega_{1} \lambda^{2}) \sin\theta_{1}\cos\theta_{2} \cdot \dot{\theta}_{1}
=0  \, .  \label{con:phi2}
\end{eqnarray}
%%%%

The above set of equations (\ref{con:psi})-(\ref{con:phi2}) implies
%%%%
\begin{equation}\label{theta:van}
\Pi_{\theta_{1}} = \dot{\theta}_{1} = 0 \, , \qquad \qquad \quad   \Pi_{\theta_{2}} = \dot{\theta}_{2} = 0 \, ,
\end{equation}
%%%%
where $\Pi_{\theta_{i}}$ are the momenta conjugate to $\theta_{i}$.

Using (\ref{theta:van}), it is now trivial to check that 
%%%%
\begin{equation}
\partial_{\tau}T_{\tau\tau} = 0 = \partial_{\tau}T_{\tau\sigma} \, ,
\end{equation}
%%%%
which satisfy the consistency requirements for the Virasoro constraints.

\subsection{The coupled pendulum model}\label{coupend}
Next, in order to study the interacting two (gravitational) pendulum system described by (\ref{eom:theta1}),
(\ref{eom:theta2}) we choose the following invariant plane in the phase space \cite{Basu:2011fw}:
%%%%
\begin{equation}\label{ps:theta}
\theta_{2} \sim 0 \, , \qquad  \Pi_{\theta_{2}} \sim \dot{\theta}_{2} \sim 0  \, .
\end{equation}
%%%%

Using (\ref{ps:theta}) the equation of motion for $\theta_{1}$ can be written as
%%%%
\begin{equation}\label{red:theta1}
\ddot{\theta}_{1} + A \sin\theta_{1} \cos\theta_{1} - \qty(B+D) \sin\theta_{1} = 0  \, ,
\end{equation}
%%%%
where 
\begin{align}
A &= \qty(1-\frac{\lambda^{2}}{\lambda_{1}^{2}}) \qty(\ell_{1}^{2} - \omega_{1}^{2}) \, ; 
\\[6pt]
B &= \frac{1}{\lambda_{1}^{2}} \left[ k\qty(\omega_{1}\ell_{3} - \alpha \ell_{1}) +
\lambda^{2} \qty(\ell_{1}\ell_{3} - \alpha \omega_{1}) \right]  \, ;  \\[6pt]
D &= \frac{1}{\lambda_{1}^{2}} \left[ k\qty(\omega_{1}\ell_{2} - \omega_{2} \ell_{1}) +
\lambda^{2} \qty(\ell_{1}\ell_{2} - \omega_{1} \omega_{2}) \right]  \, .
\end{align}
%%%%

Let $\bar{\theta}_{1}$ be the solution to (\ref{red:theta1}). In the next step, we shall consider small fluctuations ($\eta$)
about the invariant plane $\theta_{2}\sim 0$, $\Pi_{\theta_{2}} \sim 0$ which results in the normal variational equation
(NVE) of the following form:
%%%%
\begin{equation}\label{eqn:nve}
\ddot{\eta} + \qty(F - G) \eta - \tilde{D} \cos\bar{\theta}_{1} \cdot \eta = 0 \, ,
\end{equation}
%%%%
where
%%%%
\begin{align}
F &= \qty(1-\frac{\lambda^{2}}{\lambda_{2}^{2}}) \qty(\ell_{2}^{2} - \omega_{2}^{2}) \, ; 
\\[6pt]
G &= \frac{1}{\lambda_{2}^{2}} \left[ k\qty(\alpha \ell_{2} - \omega_{2}\ell_{3}) +
\lambda^{2} \qty(\ell_{2}\ell_{3} - \alpha \omega_{2}) \right]  \, ;  \\[6pt]
\tilde{D} &= \frac{1}{\lambda_{2}^{2}} \left[ k\qty(\omega_{1}\ell_{2} - \omega_{2} \ell_{1}) +
\lambda^{2} \qty(\ell_{1}\ell_{2} - \omega_{1} \omega_{2}) \right]  \, .
\end{align}
%%%%

In order to study the NVE (\ref{eqn:nve}) it will be convenient to introduce the variable $z$ such that \cite{Basu:2011fw}
%%%%
\begin{equation}\label{cng:var}
\cos\bar{\theta}_{1} = z \, .
\end{equation}
%%%%

With this change in variable, (\ref{eqn:nve}) can be expressed as
%%%%
\begin{equation}\label{Lame}
\eta''(z) + \frac{f'(z)\eta'(z)}{2 f(z)} + \qty(F - G - \tilde{D} z) \frac{\eta(z)}{f(z)} = 0 \, .
\end{equation}
%%%%

The above equation (\ref{Lame}) is similar to a second order linear homogeneous differential equation, known as
the Lam\'{e} equation \cite{Basu:2011fw}, with
%%%%
\begin{equation}\label{fz}
f(z) = \dot{\bar{\theta}}_{1}^{2} \, \sin^{2}\bar{\theta}_{1} = \left[ 2E^{2} - A \qty(1-z^{2})
- 2\qty(B+D) z \right] \qty(1-z^{2}).
\end{equation}
%%%%

It is difficult to solve the NVE (\ref{Lame}) exactly. However, we can expand the coefficients of (\ref{Lame}) for small
values of the variable $|z|$. The resulting solution can be expressed in terms of special functions and can be written as
%%%%
\begin{align}\label{etasoln}
\begin{split}
\eta(z) &= \exp \qty[ \frac{\tilde{D}z}{\Upsilon}\qty(\frac{A}{2}-1)+
\frac{z}{\Upsilon}\qty(B+D)\qty(F-G) ] \left\{ \mathcal{C}_{1} \,\,\,  
\texttt{Hermite}\qty[\frac{\chi_{1}}{8\Upsilon^{3}},\frac{\chi_{2}}{2
\Upsilon^{\frac{3}{2}}}+\frac{\Upsilon^{\frac{1}{2}}z}{A-2}]   \right.\\[4pt]
& \left. + \,\mathcal{C}_{2} \,\,\, {}_{1}\texttt{F}_{1}\qty[ -\frac{\chi_{1}}{16
\Upsilon^{3}},\frac{1}{2}, \qty(\frac{\chi_{2}}{2\Upsilon^{\frac{3}{2}}}+
\frac{\Upsilon^{\frac{1}{2}}z}{A-2})^{2} ] \right\}  \, ,
\end{split}
\end{align}
%%%%
where $\mathcal{C}_{1}$, $\mathcal{C}_{2}$ are constants of integration, 
%%%%
\begin{equation}\label{upsilon}
\Upsilon = A^{2}+\qty(B+D)^{2}-3A+2 \, ,
\end{equation}
%%%%
and the detail expressions for $\chi_{1}$ and $\chi_{2}$ are provided in appendix \ref{coefschi}. Here $\texttt{Hermite}
\qty[n,z]$ and ${}_{1}\texttt{F}_{1}\qty[a;b;z]$ are the Hermite polynomial function and Kummer confluent
hypergeometric function, respectively. Clearly, this form of the solution is not Liouvillian indicating the non-integrability
of the system \cite{Basu:2011fw,Roychowdhury:2017vdo,Nunez:2018ags,Nunez:2018qcj,Roychowdhury:2020zer,
Rigatos:2020igd}.

Furthermore, using the change of variable \cite{Roychowdhury:2017vdo,Nunez:2018ags,Nunez:2018qcj,
Roychowdhury:2020zer,Rigatos:2020igd}
%%%%
\begin{equation}
\label{schrodinger}
\eta(z) = \exp\qty[ \int \qty(w(z)-\frac{\mathcal{B}(z)}{2})\dd z ]
\end{equation}
%%%%
we can recast (\ref{Lame}) in the Schr\"{o}dinger form
%%%%
\begin{equation}\label{weqn}
w'(z) + w^{2}(z) = \mathcal{V}(z)\equiv \frac{2\mathcal{B}'(z)+\mathcal{B}^{2}(z)
-4\mathcal{A}(z)}{4} \, ,
\end{equation}
%%%%
where
%%%%
\begin{equation}\label{def:AB}
\mathcal{A}(z) = \frac{F-G-\tilde{D}z}{f(z)} \, , \qquad 
\mathcal{B}(z) = \frac{f'(z)}{2f(z)} \, .
\end{equation}
%%%%

Using (\ref{def:AB}) in (\ref{weqn}) we can compute the function $\mathcal{V}(z)$ as
%%%%
\begin{align}\label{func:V}
\begin{split}
\mathcal{V}(z) &= -\frac{3}{4\qty(z^{2}-1)^{2}} + \frac{\qty(\beta_{1}+z\tilde{\beta}_{1})}
{4\qty(z^{2}-1)}-\frac{3\beta_{2}}{4\qty(Az^{2}-2\qty(B+D)z-\qty(A-2))^{2}}      \\
&\quad +\frac{\qty(\beta_{3}+z\tilde{\beta}_{3})}{4\qty(Az^{2}-2\qty(B+D)z-\qty(A-2))}  \, ,
\end{split}
\end{align}
%%%%
where
%%%%
\begin{align}
\beta_{1} &=\frac{1}{\beta_{0}} \qty(1-A+2\tilde{D}\qty(B+D)-2\qty(F-G)) \, ; \qquad
\beta_{0} = \qty(B+D)^{2}-1  \, ,     \\[4pt]
\tilde{\beta}_{1} &= \frac{1}{\beta_{0}} \qty(\qty(1-A)\qty(B+D)-2\qty(B+D)\qty(F-G))  \, ,
\\[4pt]
\beta_{2} &= A\qty(A-2) + \qty(B+D)^{2}   \, ,   \\[4pt]
\begin{split}
\beta_{3} &= \frac{1}{\beta_{0}} \Big(A^2-2 A B^2-4 A B D-2 A B \tilde{D}-2 A D^2-2 A D
\tilde{D}+2 A F-2 A G-A  \\
\quad & -4 B^2 F+4 B^2 G+2 B^2-8 B D F+8 B D G+4 B D+4 B \tilde{D}-4 D^2 F+4 D^2 G  \\
\quad & +2 D^2+4 D \tilde{D} \Big)  \,  ,
\end{split}  \\[4pt]
\begin{split}
\tilde{\beta}_{3} &= \frac{1}{\beta_{0}} \qty(A\qty(A-1)\qty(B+D)-2A\tilde{D}+2A\qty(F-G)\qty(B+D)) \, .
\end{split}
\end{align}
%%%%

Notice that, in writing (\ref{func:V}) we have set the constant of integration $E=1$ in (\ref{fz}).

Clearly, the function $\mathcal{V}(z)$ has poles of order $2$ at
%%%%
\begin{equation}\label{posn:poles}
z=\pm1 \, , \qquad \quad z=\frac{1}{A}\qty[\qty(B+D)\pm\sqrt{A\qty(A-2)
+\qty(B+D)^{2}}] \, ,
\end{equation}
%%%%
and the order of $\mathcal{V}(z)$ at infinity can be computed to be $1$. Hence, $\mathcal{V}(z)$ does not
belong to any of the three cases discussed in the Appendix \ref{kovacic}. Therefore, the form of the solution
to equation (\ref{weqn}) must be non-Liouvillian.

In fact, for small values of $|z|$ we can series expand $\mathcal{V}(z)$ in (\ref{func:V}) and can find the
solution to (\ref{weqn}) as
%%%%
\begin{align}\label{wsol:expand}
\begin{split}
w(z) &= \frac{\chi_{3}}{\Pi_{2}} \cdot \frac{\texttt{Bi}' \qty(\Pi_{1}) + \mathcal{C}_{3} \,
\texttt{Ai}' \qty(\Pi_{1})}{\texttt{Bi} \qty(\Pi_{1})+\mathcal{C}_{3} \, \texttt{Ai} \qty(\Pi_{1})} \, ,
\end{split}
\end{align}
%%%%
where $\mathcal{C}_{3}$ is a constant of integration,
%%%%
\begin{align}\label{Pi12}
\Pi_{1} &= \frac{\chi_{3} z+\chi_{4}}{\Pi_{2}} \, ,
& \Pi_{2} &= 2^{\frac{2}{3}}\qty(A-2)\chi_{3}^{\frac{2}{3}} \, ,
\end{align}
%%%%

%%%%
\begin{align}\label{chi3}
\begin{split}
\chi_{3} &= 8 A^2 B+8 A^2 D-4 A^2 \tilde{D}-8 A B F+8 A B G-12 A B-8 A D F+8 A D G-12 A D  \\
&\quad +16 A \tilde{D}+12 B^3+36 B^2 D+36 B D^2+16 B F-16 B G-8 B+12 D^3+16 D F   \\
&\quad -16 D G-8 D-16 \tilde{D} \, ,
\end{split}
\end{align}
%%%%
%%%%
\begin{align}\label{chi4}
\begin{split}
\chi_{4} &= -4 A^3+4 A^2 F-4 A^2 G+20 A^2-3 A B^2-6 A B D-3 A D^2-16 A F+16 A G-32 A   \\
&\quad +6 B^2+12 B D+6 D^2+16 F-16 G+16 \, .
\end{split}
\end{align}
%%%%

As it is clear from the above expression (\ref{wsol:expand}), the solution to (\ref{weqn}) is written in terms of Airy
functions $\texttt{Ai}(z)$, $\texttt{Bi}(z)$ and their derivatives $\texttt{Ai}'(z)$, $\texttt{Bi}'(z)$ thereby making it
non-Liouvillian. This reassures the non-integrability of the system.

\section{Analytic integrability of the decoupled systems}\label{decoup}

In this section we shall discuss the analytic integrability of the two pendulum system in the decoupling limit. It is
easy to check from (\ref{eom:theta1}), (\ref{eom:theta2}) that the two pendula decouples from each other in the limit
%%%%
\begin{equation}\label{lim:decoup}
\frac{k}{\lambda^{2}}  =\frac{\qty(\omega_{1}
\omega_{2} - \ell_{1}\ell_{2})}{\qty(\omega_{1}\ell_{2} - \omega_{2}\ell_{1})} \, ,
\end{equation}
%%%%
and the equations of motion (\ref{eom:theta1}) and (\ref{eom:theta2}) simplify to
%%%%
\begin{align}
\ddot{\theta}_{1} + A \sin\theta_{1}\cos\theta_{1} - B\sin\theta_{1} &= 0 \, ,   \\[8pt]
\ddot{\theta}_{2} + F \sin\theta_{2}\cos\theta_{2} - G\sin\theta_{2} &= 0 \, ,
\end{align}
%%%%
where $A$, $B$, $F$ and $G$ have been defined earlier.

We notice that, when
%%%%
\begin{equation}\label{freq:wind}
\omega_{1} \leftrightarrow - \ell_{1} \, , \qquad \text{or} \qquad  \omega_{2} \leftrightarrow \ell_{2} \, ,
\end{equation}
%%%%
the decoupling limit (\ref{lim:decoup}) reduces to what was studied in \cite{Arutyunov:2020sdo,Ishii:2021asw};
namely, $\displaystyle k=\lambda^{2}$. The constraint (\ref{freq:wind}) implies that either the string must wind
anti-clockwise along the isometry direction $\phi_{1}$, or it winds in the clockwise direction along $\phi_{2}$.
Thus we may consider the limit $\displaystyle k=\lambda^{2}$ as a special case of (\ref{lim:decoup}).

Next, in order to check the integrability of the individual systems, we study their dynamics in the corresponding phase
spaces $\{\theta_{i},\Pi_{\theta_{i}}\}$ ($i=1,2$). For the first pendulum we choose the invariant plane given by
%%%%
\begin{equation}
\theta_{1}=\dot{\theta}_{1}=0 \, ,
\end{equation}
%%%%
and consider the fluctuation $\delta\theta_{1}\sim\sigma(\tau)$ around this plane to compute the NVE as
%%%%
\begin{equation}\label{NVE:1st}
\ddot{\sigma}+\qty(A-B)\sigma = 0 \, ,
\end{equation}
%%%%
which is nothing but simple harmonic motion with frequency $\sim \sqrt{A-B}$, hence trivially integrable as per differential
Galois theory \cite{Basu:2011fw}.

In a similar manner we can choose an invariant plane $\theta_{2}=\dot{\theta}_{2}=0$ in the phase space of the second
pendulum. Subsequently, the Kovacic's algorithm ensures the integrability of this system as well.

\section{Conclusions}\label{conclusion}
In this paper, we have established classical Liouvillian non-integrability of string sigma model in the recently
proposed Arutyunov-Bassi-Lacroix (ABL) background (\ref{ABL:metric}) which is a generalization of the
Einsteinian $T^{1,1}$ manifold \cite{Arutyunov:2020sdo}. We use the Kovacic's algorithm \cite{Kovacic:1986,
Saunders:1981} in order to perform our computations. Our analysis complements the claim made in
\cite{Arutyunov:2020sdo,Ishii:2021asw} and is compatible with the Lax pair formulation of
\cite{Arutyunov:2020sdo}.\medskip

We observe that, if we consider a string that winds around the deformed torus $T^{\qty(\lambda_{1},
\lambda_{2},\lambda)}_{k}$, the system can be described by two coupled harmonic oscillators. We analyse
the corresponding coupled differential equation. By appropriately choosing an invariant plane in the phase
space of the system, we analyse the corresponding normal variational equation (NVE). The solution to this
equation turns out to be non-Liouvillian which therefore establishes the non-integrability. We further recast
the NVE in the Schr\"{o}dinger form (\ref{weqn}) and carefully analyze the polynomial function (\ref{func:V})
arising from it. We observe incompatibility of this function with the Kovacic's classification (see Appendix
\ref{kovacic}), thereby establishing the non-integrability.\medskip

Next, we proceed to study the limit in which the coupled oscillators decouple and turn out to be integrable.
As a matter of fact, for the particular example of winding string, we find a \emph{generalized} decoupling
limit and, under special conditions (\ref{freq:wind}), it gives the decoupling limit studied in
\cite{Arutyunov:2020sdo,Ishii:2021asw}. This analysis reveals that, in transiting to this later limit in
\cite{Arutyunov:2020sdo,Ishii:2021asw}, the string winds (anti-)clockwise along the isometry direction
($\phi_{1}$) $\phi_{2}$.\medskip

The Kovacic's algorithm has been proven to be an excellent mathematical tool to (dis)prove analytic
integrability over general backgrounds \cite{Basu:2011b},\cite{Basu:2011fw},
\cite{Roychowdhury:2017vdo}-\cite{Rigatos:2020igd}. As has been shown in the present work, one can
reliably use this formalism as it complements the existing methodologies in the literature
\cite{Arutyunov:2020sdo,Ishii:2021asw}. It will be interesting to apply this formulation, along with the other
analytic methods, to explore integrable structures of other systems in future.

\section*{Acknowledgments}
J.P., A.M. and D.R. is indebted to the authorities of IIT Roorkee for their unconditional support towards
researches in basic sciences. A.M. acknowledges The Science and Engineering Research Board (SERB),
India for financial support. D.R. would also like to acknowledge The Royal Society, UK for financial
assistance, and acknowledges the Grant (No. SRG/2020/000088) received from The Science and
Engineering Research Board (SERB), India.
The work of A.L. is supported by the Chilean \emph{National Agency for Research and Development}
(ANID)/ FONDECYT / POSTDOCTORADO BECAS CHILE / Project No. 3190021.

\appendix
\numberwithin{equation}{section}
\renewcommand{\theequation}{\thesection\arabic{equation}}
\section{The Kovacic's algorithm}\label{kovacic}

The Kovacic's algorithm is a systematic procedure to check the Liouvillian non-integrability of a dynamical
system \cite{Kovacic:1986,Saunders:1981}. This algorithm is implemented in order to realize whether the
second order linear homogeneous differential equation of the type
%%%%
\begin{equation}\label{ode:kova}
\eta''(z)+\mathcal{M}(z)\eta'(z)+\mathcal{N}(z)\eta(z)=0
\end{equation}
%%%%
with polynomial coefficients $\mathcal{M}(z)$, $\mathcal{N}(z)$ are integrable in the Liouvillian sense. We
look for solutions of (\ref{ode:kova}) in the Liouvillian form; namely, those solutions which can be expressed
in terms of simple algebraic polynomial functions, exponentials, or trigonometric functions. As a matter of
fact, this algorithm is a variant of the Galois theory of differential equations (the Piccard-Vessoit theory
\cite{Kovacic:2005}) and utilizes the $SL\qty(2,\mathbb{C})$ group of symmetries of the differential equation
(\ref{ode:kova}) in its analysis.\smallskip

Without going into detailed mathematical analysis which is rather involved, we limit ourselves to find the
relation between the coefficients $\mathcal{M}(z)$, $\mathcal{M}'(z)$ and $\mathcal{N}(z)$ that makes
(\ref{ode:kova}) integrable. In order to do so, we perform the following change of variable:
%%%%
\begin{equation}
\label{var:cng}
\eta(z) = \exp\qty[ \int \qty(w(z)-\frac{\mathcal{M}(z)}{2})\dd z ].
\end{equation}
%%%%

This allows us to rewrite (\ref{ode:kova}) as
%%%%
%%%%
\begin{equation}\label{wkova}
w'(z) + w^{2}(z) = \mathcal{V}(z)\equiv \frac{\mathcal{M}'(z)+\mathcal{M}^{2}(z)
-4\mathcal{N}(z)}{4} \, ,
\end{equation}
%%%%
where `$\,' \,$' denotes differentiation w.r.to $z$.\smallskip

The group of symmetry transformations of the solutions of (\ref{ode:kova}) is in fact a subgroup $G$ of $SL
\qty(2,\mathbb{C})$ which can be classified as follows: (i) For any $ a, \, b \in \mathbb{C}$, if $G$
is generated by the matrix $M = \qty( \{a, \, 0 \}, \{ b, \, 1/a \} )$ then $w(z)$ is a rational polynomial function
of degree $1$; (ii) When $M = \qty( \{a, \, 0 \}, \{ 0, \, 1/a \} )$, $M = \qty( \{0, \, a \}, \{ -1/a, \, 0 \} )$ $w(z)$
is a function of degree $2$; (iii) When $G$ is a finite group and not generated by a matrix of the form
mentioned in (i) and (ii) above, $w(z)$ is a function of degree $4$, $6$, or $12$; (iv) When $G=SL\qty(2,
\mathbb{C})$, any non-vanishing solution $w(z)$ is not Liouvillian.\smallskip

The Kovacic's algorithm also proposes a set of three necessary but not sufficient conditions for the rational
polynomial function $\mathcal{V}(z)$ that must be compatible with the aforementioned group theoretical
analysis \cite{Kovacic:1986}. These are as follows: (i) $\mathcal{V}(z)$ has pole of order $1$, or $2n$
($n\in\mathbb{Z}^{+}$). Also, the order of $\mathcal{V}(z)$ at infinity, defined as the highest power of the
denominator minus that of the numerator, is either $2n$ or greater than 2; (ii) $\mathcal{V}(z)$ either has
pole of order $2$, or poles of order $2n+1$ greater than $2$; (iii) $\mathcal{V}(z)$ has poles not greater
than $2$ and the order of $\mathcal{V}(z)$ at infinity is at least $2$. If none of these conditions are met, the
solution to (\ref{ode:kova}) is non-Liouvillian and ensures the non-integrability of (\ref{ode:kova}). On the
other hand, fulfilment of any one of the above conditions makes us eligible to apply the Kovacic's algorithm
to the ODE (\ref{ode:kova}). It is then necessary to determine whether $w(z)$ is a polynomial function of
degree $1$, $2$, $4$, $6$, or $12$ in which case (\ref{ode:kova}) is integrable.

\section{Expressions for the constants $\chi_{1}$ and $\chi_{2}$ in (\ref{etasoln})}\label{coefschi}

%%%%
\begin{align}\label{chi1}
\begin{split}
\chi_{1} &= -4 F A^5+4 G A^5+\tilde{D}^2 A^4+2 B \tilde{D} A^4+2 D \tilde{D} A^4+32 F A^4-32 G A^4
-8 \tilde{D}^2 A^3-14 B \tilde{D} A^3   \\
& \quad -14 D \tilde{D} A^3-4 B^2 F A^3-4 D^2 F A^3-8 B D F A^3+4 B \tilde{D} F A^3+4 D \tilde{D} F A^3
-100 F A^3+4 B^2 G A^3   \\
& \quad +4 D^2 G A^3+8 B D G A^3-4 B \tilde{D} G A^3-4 D \tilde{D} G A^3+100 G A^3+24 \tilde{D}^2 A^2
+4 B^2 F^2 A^2+4 D^2 F^2 A^2   \\
&\quad +8 B D F^2 A^2+4 B^2 G^2 A^2+4 D^2 G^2 A^2+8 B D G^2 A^2+2 B^3 \tilde{D} A^2+2 D^3 \tilde{D} A^2
+6 B D^2 \tilde{D} A^2  \\
&\quad +36 B \tilde{D} A^2+6 B^2 D \tilde{D} A^2+36 D \tilde{D} A^2+20 B^2 F A^2+20 D^2 F A^2+40 B D F A^2
-24 B \tilde{D} F A^2  \\
&\quad -24 D \tilde{D} F A^2+152 F A^2-20 B^2 G A^2-20 D^2 G A^2-40 B D G A^2+24 B \tilde{D} G A^2
+24 D \tilde{D} G A^2    \\
&\quad -8 B^2 F G A^2-8 D^2 F G A^2-16 B D F G A^2-152 G A^2-32 \tilde{D}^2 A-16 B^2 F^2 A-16 D^2 F^2 A
    \\
&\quad -32 B D F^2 A -16 B^2 G^2 A-16 D^2 G^2 A-32 B D G^2 A-8 B^3 \tilde{D} A-8 D^3 \tilde{D} A-24 B D^2 \tilde{D} A
    \\
&\quad -40 B \tilde{D} A-24 B^2 D \tilde{D} A-40 D \tilde{D} A-32 B^2 F A-32 D^2 F A-64 B D F A+48 B \tilde{D} F A
    \\
&\quad +48 D \tilde{D} F A -112 F A+32 B^2 G A+32 D^2 G A+64 B D G A-48 B \tilde{D} G A-48 D \tilde{D} G A
+32 B^2 F G A    \\
&\quad +32 D^2 F G A+64 B D F G A+112 G A+16 \tilde{D}^2+16 B^2 F^2+16 D^2 F^2+32 B D F^2+16 B^2 G^2
+16 D^2 G^2   \\
&\quad+32 B D G^2+8 B^3 +32 B D G^2+8 D^3 \tilde{D}+24 B D^2 \tilde{D}+16 B \tilde{D}+24 B^2 D \tilde{D}
+16 D \tilde{D}    \\
&\quad +16 B^2 F+16 D^2 F+32 B D F-32 B \tilde{D} F-32 D \tilde{D} F+32 F-16 B^2 G-16 D^2 G-32 B D G
+32 B \tilde{D} G   \\
&\quad +32 D \tilde{D} G-32 B^2 F G-32 D^2 F G-64 B D F G-32 G .
\end{split}
\end{align}
%%%%

%%%%
\begin{align}\label{chi2}
\begin{split}
\chi_{2} &= -A^2 B-A^2 D-A^2 \tilde{D}-2 A B F+2 A B G+3 A B-2 A D F+2 A D G+3 A D   \\
&\quad +4 A \tilde{D}-B^3-3 B^2 D-3 B D^2+4 B F-4 B G-2 B-D^3+4 D F-4 D G-2 D-4 \tilde{D} .
\end{split}
\end{align}
%%%%

%%%%%%%%%%

\end{document}